\documentclass[preprint,aps,floatfix,nofootinbib,superscriptaddress]{revtex4}

\usepackage{graphicx}
\usepackage{amsmath}
\usepackage{amssymb}
\usepackage{longtable}

\newcommand{\rd}{\partial}

\newcommand{\la}{\langle}
\newcommand{\ra}{\rangle}
\newcommand{\NssN}{\la N|\bar{s}s|N\ra}
\newcommand{\NllN}{\la N|\bar{u}u+\bar{d}d|N\ra}

\newcommand{\dt}{\Delta t}
\newcommand{\dts}{{\Delta t}_s}
\newcommand{\1}{\mbox{1}\hspace{-0.25em}\mbox{l}}
\newcommand{\be}{\begin{equation}}
\newcommand{\ee}{\end{equation}}
\newcommand{\bea}{\begin{eqnarray}}
\newcommand{\eea}{\end{eqnarray}}
\newcommand{\bi}{\begin{itemize}}
\newcommand{\ei}{\end{itemize}}
\newcommand{\bfy}{{\bf y}}
\newcommand{\Os}{\mathcal O_S^{\rm lat}}
\newcommand{\NslatN}{\la N|\Os|N\ra}

\begin{document}

\vspace*{-10mm}
\begin{flushright}
\normalsize
KEK-CP-276        \\
\end{flushright}

\title{
Nucleon strange quark content
from $N_f\!=\!2+1$ lattice QCD with exact chiral symmetry
}

\newcommand{\KEK}{
  KEK Theory Center,
  High Energy Accelerator Research Organization (KEK),
  Tsukuba 305-0801, Japan
}
\newcommand{\CRC}{
  Computing Research Center,
  High Energy Accelerator Research Organization (KEK),
  Tsukuba 305-0801, Japan
}
\newcommand{\Tsukuba}{
  Graduate School of Pure and Applied Sciences, University of Tsukuba,
  Tsukuba, Ibaraki 305-8571, Japan
}
\newcommand{\CCS}{
  Center for Computational Sciences, University of Tsukuba, Tsukuba, 
  Ibaraki 305-8577, Japan
}
\newcommand{\GUAS}{
  School of High Energy Accelerator Science,
  The Graduate University for Advanced Studies (Sokendai),
  Tsukuba 305-0801, Japan
}
\newcommand{\Osaka}{
  Department of Physics, Osaka University,
  Toyonaka 560-0043, Japan
}
\newcommand{\Nagoya}{
  Kobayashi-Maskawa Institute for the Origin of Particles and the
  Universe~(KMI), 
  Nagoya University, Nagoya, Aichi 464-8602, Japan
}

\author{H.~Ohki}
\affiliation{\Nagoya}

\author{K.~Takeda}
\affiliation{\KEK}

\author{S.~Aoki}
\affiliation{\Tsukuba}
\affiliation{\CCS}

\author{S.~Hashimoto}
\affiliation{\KEK}
\affiliation{\GUAS}

\author{T.~Kaneko}
\affiliation{\KEK}
\affiliation{\GUAS}

\author{H.~Matsufuru}
\affiliation{\CRC}

\author{J.~Noaki}
\affiliation{\KEK}

\author{T.~Onogi}
\affiliation{\Osaka}

\collaboration{JLQCD Collaboration}
\noaffiliation

\date{\today}

\begin{abstract}

We calculate the strange quark content of the nucleon $\NssN$ 
in $2+1$ \!-flavor lattice QCD. 
Chirally symmetric overlap fermion formulation is used
to avoid the contamination from up and down quark contents
due to an operator mixing between strange and light scalar operators,
$\bar{s}s$ and $\bar{u}u+\bar{d}d$.
At a lattice spacing $a=0.112(1)$~fm,
we perform calculations at four values of degenerate up and down quark
masses $m_{ud}$, 
which cover a range of the pion mass $M_\pi \!\simeq$~300\,--\,540~MeV. 
We employ two different methods to calculate $\NssN$.
One is a direct method where we calculate $\NssN$ by directly inserting
 the $\bar ss$ operator. The other is an indirect method where $\NssN$
 is extracted  from a derivative of the nucleon mass in terms of the
 strange quark mass.
With these two methods we obtain consistent results for $\NssN$
with each other.
Our best estimate 
$f_{T_s}=m_s\NssN/M_N=0.009(15)_{\rm stat}(16)_{\rm sys}$
is in good agreement with our previous studies in two-flavor QCD.

\end{abstract}

\pacs{}

\maketitle

\section{Introduction}

The bulk of the nucleon mass $M_N$ is produced by dynamically broken
chiral symmetry in the vacuum of Quantum Chromodynamics (QCD).
This should happen even in the limit of vanishing up and down
(current) quark masses.
Yet, there are also contributions from non-zero bare masses of up,
down and strange quarks, that are given by a matrix element
$m_q\langle N|\bar{q}q|N\rangle$
of a scalar operator $\bar{q}q$ made of quark field $q$ with mass
$m_q$ evaluated on the nucleon state $|N\rangle$.
This quantity is of fundamental importance to characterize the nucleon
structure.
More recently, this quantity, especially that of strange quark, is
attracting further interest as it determines the cross section of
possible dark matter particles to hit the nucleus and 
thus to determine the sensitivity of dark matter search experiments
(see, for instance, \cite{L_eff}).

The fraction of nucleon mass made of non-vanishing quark masses is
conveniently parametrized as
\begin{equation}
  f_{T_q} = \frac{m_q\langle N|\bar{q}q|N\rangle}{M_N}.
\end{equation}
The light quark contents $f_{T_{\{u,d\}}}$ can be related to the 
$\pi N$ sigma term $\sigma_{\pi N}$, which is determined from
experimental data of the $\pi N$ scattering amplitude.
Evaluation of the strange quark content $f_{T_s}$ is more involved.
One uses $\sigma_{\pi N}$ and a phenomenological estimate of the
flavor SU(3) violation parameter
$\sigma_0=m_{ud}\langle N|\bar{u}u+\bar{d}d-2\bar{s}s|N\rangle$,
where $m_{ud}$ is (degenerate) up and down quark mass.
Recent experimental data $\sigma_{\pi N}$ = 64(7)~MeV \cite{piN:exprt}
and $\sigma_0$ = 36(7)~MeV obtained from heavy baryon chiral
perturbation theory (HBChPT) \cite{sigma_0:hbchpt} led to $f_{T_s}$ = 0.41(9).
This large value appeared to be puzzling, as it suggests that the
strange quark plays major role to construct nucleon.
Early lattice calculations
\cite{Fukugita:1994ba,Dong:1995ec,Gusken:1998wy} also suggested such
large value. 

\begin{figure}[tbp]
 \centering
 \includegraphics[width=0.3\textwidth,clip]{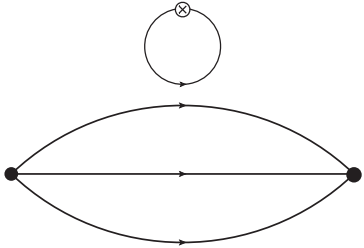}
 \caption{
 Nucleon three-point function used to determine $\NssN$.
 Solid lines represent quark propagators that are dressed by 
 gluons and sea quarks. 
 Connected three lines form the nucleon propagator,
 whereas the disconnected quark loop 
 arises from the strange scalar operator $\bar{s}s$.
 }
 \label{fig:diagram}
\end{figure}

In our previous studies~\cite{Ohki:2008ff,Takeda:2010cw},
we carried out non-perturbative calculations of $\NssN$ in two-flavor QCD, 
where up and down quarks are assumed to be degenerate.
In Ref.~\cite{Ohki:2008ff},
$\NssN$ is indirectly estimated from the $m_s$ dependence of $M_N$
through the Feynman-Hellmann theorem
\begin{equation}
 \frac{\rd M_N}{\rd m_{s}}=\langle N|\bar{s}s|N\rangle.
  \label{eqn:FH}
\end{equation}
We refer to this method as the spectrum method in this paper. 
In Ref.~\cite{Takeda:2010cw}, on the other hand, 
$\NssN$ is extracted directly from a disconnected three-point function
of the nucleon (Fig.~\ref{fig:diagram}).
Since we use a ratio of the three- and two-point functions 
(see (\ref{eqn:ratio}) in Sec.~\ref{Sec:result})
to improve the accuracy of $\NssN$,
this method is referred to as the ratio method in the following.
These two studies consistently yielded $f_{T_s}\!\lesssim\!0.05$
which is significantly smaller than the phenomenological estimate.

In this paper,
we extend our previous studies to $2+1$ \!-flavor QCD.
This is a necessary step towards a realistic calculation of $\NssN$,
since effects of dynamical strange quarks are 
difficult to estimate analytically.
In addition, 
we can eliminate a subtlety in the spectrum method when used for
two-flavor QCD. Namely, since this theory does not have strange sea quark,
we estimated $\NssN$ as a derivative in terms of up and down sea quark mass
at sea ($m_{ud, \rm sea}$) and valence ($m_{ud, \rm val}$) quark masses
set to the physical strange quark mass $m_{s, \rm phys}$
\begin{equation}
  \langle N|\bar{s}s|N\rangle  
  = \left. 
       \frac{\rd M_N}{\rd m_{ud, \rm sea}} 
    \right|_{m_{ud, \rm sea}=m_{ud, \rm val}=m_{s, \rm phys}},
  \label{eqn:FH:Nf2}
\end{equation}
assuming that $\NssN$ mildly depends on the quark masses.
This assumption is eliminated in the present work.

A number of lattice studies of the strange quark content
have been recently performed 
in $N_f\!=\!2$~\cite{Michael:2001bv,Bali:2011ks,Babich:2010at},
$2\!+\!1$~\cite{Young:2009zb,Toussaint:2009pz,Engelhardt:2010zr,Gong:2012nw,Durr:2011mp,Horsley:2011wr,Freeman:2012ry,Shanahan:2012},
and $2\!+\!1\!+\!1$~\cite{Freeman:2012ry,Dinter:2012tt} QCD
using either the spectrum~\cite{Michael:2001bv,Young:2009zb,Durr:2011mp,Horsley:2011wr,Shanahan:2012} or ratio~\cite{Bali:2011ks,Babich:2010at,Toussaint:2009pz,Engelhardt:2010zr,Gong:2012nw,Freeman:2012ry,Dinter:2012tt} method.
An important advantage of our work over the previous calculations
is that chiral symmetry is preserved
by employing the overlap quark action~\cite{Neuberger:1997fp,Neuberger:1998wv}.
Conventional Wilson-type fermions, which explicitly violate
chiral symmetry at finite lattice spacings, induce a
mixing of scalar operators between
$\bar{s}s$ and $\bar{u}u+\bar{d}d$~\cite{Takeda:2010cw}.
The nucleon three-point function in Fig.~\ref{fig:diagram} then
receives a contribution from a connected diagram 
with the $\bar{u}u+\bar{d}d$ operator through the renormalization of $\bar{s}s$.
The connected contribution is larger than the disconnected one typically
by an order of magnitude,
and a subtraction of such a large contamination gives rise to
a substantial uncertainty in $\NssN$ \cite{Michael:2001bv}. This serious
problem is entirely avoided in our work using the chiral lattice fermion
formulation. 

This paper is organized as follows. 
We describe our simulation setup to generate gauge ensembles and 
to calculate relevant nucleon correlators in Sec.~\ref{Sec:setup}.
The strange quark content is extracted 
through the ratio and spectrum methods at simulated quark masses 
in Sec.~\ref{Sec:result}.
We then extrapolate these results to the physical point 
in Sec.~\ref{Sec:Chiral}.
Our conclusions are given in Sec.~\ref{Sec:Conclusion}.  
Our preliminary reports of this work are found 
in Refs.~\cite{Ohki:lat09,Takeda:lat10}.

\section{Simulation method}
\label{Sec:setup}

\subsection{Gauge ensembles}

We simulate QCD with degenerate up and down quarks and heavier  
strange quarks.
Chiral symmetry is exactly preserved by employing 
the overlap quark action~\cite{Neuberger:1997fp,Neuberger:1998wv}.
Its Dirac operator is given by 
\begin{equation}
 D(m)
  =
  \left( m_0+\frac{m}{2} \right)
  +
  \left( m_0-\frac{m}{2} \right) \gamma_5 \, {\rm sgn}[H_W(m_0)],
  \label{overlap}
\end{equation}
where $m$ is the quark mass and 
$H_W=\gamma_5 D_W$ is the Hermitian Wilson-Dirac operator.
The mass parameter of $H_W$ is chosen as $m_0\!=\!-1.6$
so that the overlap-Dirac operator $D(m)$ 
has good locality~\cite{Nf2:Conf:JLQCD}.
For gauge fields, 
we use the Iwasaki action~\cite{Iwasaki:2011jk} 
with a modification proposed in Ref.~\cite{Fukaya:2006vs}.
This leads to an extra Boltzmann factor
$\det[ H_W^2 ] /\det[ H_W^2 + \mu^2]$ ($\mu\!=\!0.2)$ 
which does not change the continuum limit of the theory
but remarkably reduces the computational cost
to calculate $\mbox{sgn}[ H_W]$ in (\ref{overlap}) 
by suppressing (near-)zero modes of $H_W$.
This Boltzmann factor prohibits
tunnelings among different topological sectors,
and we simulate only trivial topological sector in this study.
The effect of fixing topology is suppressed 
by inverse power of the lattice volume~\cite{Aoki:2007ka}
and turned out to be small, typically 1\,\% level,
in our previous studies~\cite{Noaki:lat10,m_s}. 
This small effect can be safely neglected  
with our statistical accuracy for baryon observables.

Our gauge ensembles are generated at a gauge coupling $\beta=2.3$,
where the lattice spacing is determined as $a=0.112(1)$~fm 
using the $\Omega$ baryon mass as input.
On a $N_s^3 \!\times\! N_t \!=\! 16^3\times 48$ lattice, 
we simulate two values of the degenerate up and down quark masses
$m_{ud}\!=\!0.035$ and 0.050,
and two strange quark masses $m_s\!=\!0.080$ and 0.100. 
Their physical values $m_{ud,phys}\!=\!0.0029$ and $m_{s,phys}\!=\!0.081$ 
are fixed by using $M_\pi$ and $M_K$ as inputs~\cite{m_s}.
Note also that we quote bare values in lattice units for these quark masses.
We push our simulations to two smaller $m_{ud}$'s,
0.015 and 0.025, 
on a larger lattice $24^3\times 48$
at a single value of $m_s\!=\!0.080$, which is very close to $m_{s,phys}$. 

\begin{table}[tbp]
 \begin{center}
  \begin{tabular}{l|ccc|ccc|cc|cc} \hline\hline
   $m_{ud}$ & 0.015 & & & 0.025 & & & 0.035 & & 0.050  \\ 
   $m_s$ &  0.080 & 0.080 & 0.100 & 0.080 & 0.080 & 0.100 & 0.080 &
				   0.100 & 0.080 & 0.100 \\ 
   $L/a$ & 24 & 16 & 16 & 24 & 16 & 16 & 16 & 16 & 16 & 16\\
   \hline\hline
  \end{tabular}
 \end{center}
 \caption{
 Summary of parameters used in the lattice simulation.
 } 
 \label{tab:mq_and_L}
\end{table}

Four values of $m_{ud}$ cover a range of the pion mass 
$M_\pi \simeq 300$\,--\,540~MeV.
The spatial extent $L$ is chosen 
to satisfy a condition $M_\pi L \gtrsim 4$ 
to control finite volume effects.
We carry out additional simulations 
at the two smallest $m_{ud}$'s 
on the smaller lattice size $16^3 \!\times\! 48$
to directly examine the finite volume effects.
Our simulation parameters are summarized in Table~\ref{tab:mq_and_L}.

The statistical samples at each simulation point $(m_{ud},m_s,L)$
consist of 2,500 hybrid Monte Carlo trajectories, out of which we use 500 and 50
to calculate the correlation functions in the spectrum and ratio
methods, respectively.
We employ the jackknife method with a bin size of 50 trajectories
to estimate statistical errors of the nucleon correlators
and any quantities determined from them.

On these gauge ensembles, 
we calculate the two-point nucleon correlation function
using an interpolating operator $N=\epsilon^{abc}(u_a^TC\gamma_5d_b)u_c$
with $C=\gamma_4\gamma_2$. After taking contractions, we obtain
\begin{eqnarray}
   \langle C_{2\rm pt}(\bfy,t,\dt) \rangle
   & = & 
  - \frac{1}{2N_s^3} 
   \sum_{\Gamma=(1\pm\gamma_4)/2}
   \sum_{\bf x} \epsilon^{abc}\epsilon^{a'b'c'} \notag \\
 &&  \bigg\langle \,
   \mathrm{tr}_s  [ \Gamma (D^{-1}(m))^{aa'} ]
   \mathrm{tr}_s  [ \Gamma (D^{-1}(m))^{bb'}
   (C\gamma_5)((D^{-1}(m))^{cc'})^T (C\gamma_5) ] \notag\\
   &&\qquad 
    +\mathrm{tr}_s  [ \Gamma (D^{-1}(m))^{aa'}(C\gamma_5)
    ((D^{-1}(m))^{cc'})^T
    (C\gamma_5) (D^{-1}(m))^{bb'} ] \,\bigg\rangle,
   \label{eqn:2pt}
\end{eqnarray}
where the trace ``${\rm tr_s}$'' is over spinor indices 
and $\langle \cdots \rangle$ represents a Monte Carlo average. 
Here, the quark propagators $D^{-1}(m)$ propagate from $({\bf y},t)$ to
$({\bf x},t+\dt)$. 
In order to improve statistical accuracy,
$C_{2\rm pt}$ is averaged over two choices of the projector 
$\Gamma=(1\pm \gamma_4)/2$, 
which correspond to the forward and backward propagating nucleons,
respectively. Here and in the following, for $\Gamma=(1-\gamma_4)/2$, $\dt$
is taken as $-\dt$.

We also calculate the three-point function with
a scalar operator on the lattice defined as 
\begin{equation}
  \Os=\bar{s}\biggl(1-\frac{D(0)}{2m_0}\biggr)s 
   \label{eqn:S_lat}
\end{equation}
to respect chiral symmetry in the continuum limit.

\subsection{All-to-all propagator}

As shown in Fig.~\ref{fig:diagram},
the three-point function $C_{3\rm pt}$ on a given gauge configuration
can be decomposed into two pieces.
Namely, we can write $C_{3\rm pt}$ as
\bea
   \langle C_{3{\rm pt}}(\bfy,t,\dt,\dts)\rangle
   & = &
   \langle C_{2{\rm pt}}(\bfy,t,\dt) \, 
   S^{\rm lat}(t+\dts)\rangle,
   \label{eqn:3pt_pieces}
\eea
where 
$C_{2{\rm pt}}(\bfy,t,\dt)$ is the two-point function and 
\begin{eqnarray}
 S^{\rm lat}(t+\dts)
  = 
  \frac{1}{N_s^3} \sum_{\bf z}
  \left\{ 
   \mathrm{Tr}(D^{-1}(m))(z, z)|_{z_0=t+\dts} 
   - \bigg\langle
  \mathrm{Tr}(D^{-1}(m))(z, z)|_{z_0=t+\dts}
  \bigg\rangle \,
  \right\},
   \label{eqn:3pt_loop}
\end{eqnarray}
is the scalar quark loop calculated on this configuration. The trace
``{\rm Tr}'' is over both spinor and color indices.
The nucleon piece $C_{2{\rm pt}}$ can be calculated
by using the conventional ``point-to-all'' quark propagator
$D^{-1}(x,x^\prime)$, 
the source point of which ($x^\prime$) 
has to be fixed to a certain lattice site.
The calculation of the quark-loop pieces $S^{\rm lat}$
is computationally more demanding, as it involves quark loops starting
from arbitrary lattice sites $({\bf z},t+\dts)$.
We therefore employ
the ``all-to-all'' quark propagator~\cite{Bali:2005fu,Foley:2005ac}
that contains the quark propagating from any lattice site to any site.

Let us consider a decomposition of the quark propagator 
to the contribution from low-lying eigenmodes of the Dirac operator $D(m)$ 
and that from the remaining modes
\bea
 D^{-1}(m)
 =
 \{D^{-1}(m)\}_{\rm low} +  \{D^{-1}(m)\}_{\rm high}.
 \label{eqn:Dinv}
\eea
It is expected that the low-mode contribution $\{D^{-1}(m)\}_{\rm low}$
dominates low-energy observables in QCD including $\NssN$.
We calculate it exactly as
\begin{equation}
   \{D^{-1}(m)\}_{\rm low}(x,y)
   =
   \sum_{k=1}^{N_e}\frac{1}{\lambda_k(m)}v_k(x)v_k(y)^{ \dagger},
   \label{eqn:DovL}
\end{equation}
where $\lambda_k(m)$ and $v_k(x)$ are the $k$-th lowest eigenvalue and
its associated eigenvector of $D(m)$,
and $N_e$ is the number of the low-lying modes
prepared for this calculation.

The small contribution from the remaining high-modes 
is calculated stochastically by the noise method~\cite{Dong:1993pk}.
We generate a single complex $Z_2$ noise vector $\eta(x)$ for each
configuration and split it into $N_{d}=3\times 4\times N_t/2$ vectors
$\eta^{(d)}(x)$ ($d=1,...,N_d$), 
which have nonzero elements only for a single combination of color and
spinor indices on two consecutive time slices. 
For each ``split'' noise vector $\eta^{(d)}$,
we solve a linear equation 
\begin{equation}
  \{D(m) \psi^{(d)}\}(x) = ({\mathcal P}_{\rm high}\eta^{(d)})(x) \qquad
  (d=1,...,N_d),
\end{equation} 
where ${\mathcal P}_{\rm high}\!=\!1-{\mathcal P}_{\rm low}$
and ${\mathcal P}_{\rm low}$ 
is the projector to the subspace spanned by the low-modes
\bea
 {\mathcal P}_{\rm low}(x,y)
 = 
 \sum_{k=1}^{N_e} v_k(x) v_k(y)^\dagger.
\eea
The high-mode contribution is then estimated as 
\begin{equation}
   \{D^{-1}(m)\}_{\rm high}(x,y)
   =
   \sum_{d=1}^{N_{d}}\psi^{(d)}(x)\eta^{(d)}(y)^{\dagger}.
   \label{eqn:DovH}
\end{equation}

We calculate the low- and high-mode contributions to $S^{\rm lat}$ as 
\bea
   S^{\rm lat}(t+\dts)
   & = & 
   S_{\rm low}^{\rm lat}(t+\dts)
 + S_{\rm high}^{\rm lat}(t+\dts),
   \label{eqn:3pt_loop_LH}
\eea
with 
\bea
   S_{\rm low(high)}^{\rm lat}(t+\dts)
   & = & 
   \frac{1}{N_s^3} \sum_{\bf z}
   \left.
      \{D^{-1}(m)\}_{\rm low(high)}(z,z)
   \right|_{z_0=t+\dts},
\eea
where the subtraction of the vacuum expectation value 
is assumed though it is not
written explicitly for notational simplicity.

\subsection{Low-mode averaging (LMA)}

The low-lying modes of $D(m)$ are also useful to 
precisely calculate the nucleon piece $C_{2{\rm pt}}$
in both $C_{2\rm pt}$ and $C_{3\rm pt}$. 
By applying (\ref{eqn:Dinv}),
we can decompose $C_{2{\rm pt}}$ into the following eight contributions 
\bea
   C_{{\rm 2pt}} 
   =
   C_{{\rm 2pt}}^{lll} + C_{{\rm 2pt}}^{llh} + C_{{\rm 2pt}}^{lhl}
 + C_{{\rm 2pt}}^{hll} + C_{{\rm 2pt}}^{lhh} + C_{{\rm 2pt}}^{hlh}
 + C_{{\rm 2pt}}^{hhl} + C_{{\rm 2pt}}^{hhh}.
   \label{eqn:2pt:contri}
\eea
Here, $C_{\rm 2pt}^{lll}$ is constructed only from $\{D^{-1}(m)\}_{\rm low}$.
For $C_{\rm 2pt}^{llh}$, $\{D^{-1}(m)\}_{\rm low}$ is used for two of
the valence quark propagators 
and $\{D^{-1}(m)\}_{\rm high}$ for the remaining one.
The other combinations are understood in a similar manner.
In principle, we can use the all-to-all propagator,
(\ref{eqn:DovL}) and (\ref{eqn:DovH}), 
to calculate these contributions. 
These quantities however decay exponentially 
with a large nucleon mass $M_N$ 
as the temporal separation $\dt$ increases.
At large separations, the high-mode contributions, such as $C_{{\rm
2pt}}^{hhh}$, are not sufficiently precise
with $\{D^{-1}(m)\}_{\rm high}$ evaluated 
using only single noise sample for each configuration.

We therefore use the low-mode averaging~(LMA) 
technique~\cite{DeGrand:2004qw,Giusti:2004yp} in this study.
 The low-mode part of the all-to-all propagator (\ref{eqn:DovL})
is used to calculate $C_{{\rm 2pt}}^{lll}$, which dominantly contributes 
to the nucleon correlators $C_{2\rm pt}$ and $C_{3\rm pt}$. 
We then take average of $C_{{\rm 2pt}}^{lll}(\bfy,t,\dt)$ 
over the location of the nucleon source operator $(\bfy,t)$
to largely reduce its statistical fluctuation.

The remaining and small contributions 
$\{C_{{\rm 2pt}}^{llh},...,C_{{\rm 2pt}}^{hhh}\}$
are calculated using the point-to-all quark propagator after
projecting by ${\mathcal P}_{\rm low}$ and $1\!-\!{\mathcal P}_{\rm
low}$ for $l$ and $h$ pieces, respectively.
We improve the statistical signal of these contributions 
by averaging over $(\bfy,t)$.
In order to reduce the computational cost of 
the re-calculation of the point-to-all propagators,
these contributions are averaged over a limited set of $(\bfy,t)$ 
compared to that for $C_{{\rm 2pt}}^{lll}$.

The sets of the source point
as well as the number of the low-modes $N_e$ 
are chosen differently 
for our calculations with the ratio and spectrum methods,
because the latter uses $C_{2\rm pt}$ calculated 
in the course of our study of the light meson spectrum~\cite{Noaki:lat10,m_s}. 
We summarize our choices for these two methods in the following subsections.

\subsection{Setup for the ratio method}
\label{subsec:meas:ratio}

We use $N_e\!=\!160$ and 240 low-lying modes 
on the $16^3 \! \times \! 48$ and $24^3 \! \times \! 48$ lattices,
respectively, to calculate low-mode contribution
$S_{\rm low}^{\rm lat}(t+\dt)$ and $C_{{\rm 2pt}}^{lll}(\bfy,t,\dt)$
in the ratio method.
As mentioned above,
the latter is averaged over 16 spatial points 
\bea
   \bfy
   & \in & 
   \left\{
      (0,0,0), (0,0,N_s/2), (0,N_s/2,N_s/2),
      (N_s/2,N_s/2,N_s/2), (N_s/4,N_s/4,N_s/4), 
   \right.
   \notag \\ 
   &&
   \left.
      (N_s/4,N_s/4,3N_s/4), (N_s/4,3N_s/4,3N_s/4), (3N_s/4,3N_s/4,3N_s/4),
   \right.
   \notag \\ 
   &&
   \left.
      \mbox{ and their permutations}
   \right\}
   \label{label:lma:spatial:ratio}
\eea
at each time slice $t$. 
Averaging over more points does not help to further reduce 
the statistical fluctuation of $C_{2\rm pt}$ and $C_{3\rm pt}$
because of the correlation among $C_{{\rm 2pt}}^{lll}(\bfy,t,\dt)$ 
at different spatial points $\bfy$'s.
We average $\{C_{{\rm 2pt}}^{llh},...,C_{{\rm 2pt}}^{hhh}\}$
over four time slices $t\!=\!0$, 12, 24 and 36
with the spatial location $\bfy$ kept fixed.

At heaviest $m_{ud}\, (\!=\!0.050$),
we slightly modify the setup of LMA to calculate $C_{3\rm pt}$.
With (\ref{eqn:3pt_loop_LH}) and (\ref{eqn:2pt:contri}), 
$C_{{3\rm pt}}$ on a given configuration can be rewritten as 
\bea
   C_{3\rm pt}
   & = &
   C_{{\rm 2pt}}^{lll}\,S_{\rm low}^{\rm lat}
  +\left\{C_{{\rm 2pt}}^{llh}+\cdots+C_{{\rm 2pt}}^{hhh}\right\}\,
   S_{\rm low}^{\rm lat}
   \notag \\
   &&
  +C_{{\rm 2pt}}^{lll}\,S_{\rm high}^{\rm lat}
  +\left\{C_{{\rm 2pt}}^{llh}+\cdots+C_{{\rm 2pt}}^{hhh}\right\}\,
   S_{\rm high}^{\rm lat}.
\eea
The first term represents the low-mode contribution, which gives a
dominant contribution to $C_{\rm 3pt}$ especially for small $m_{ud}$. 
Other three terms are relatively minor contributions, and their
statistical fluctuation is not substantially reduced by LMA.
At the largest $m_{ud}$, the statistical error becomes even larger
when LMA is used. We therefore apply LMA only for the first term in
that case.

The above setup of LMA typically leads to a factor of 4 (7) reduction 
of the statistical error of $C_{2\rm pt}$ ($C_{3\rm pt}$)
at our simulated values of $m_{ud}$ and $m_s$.

In our previous study in two-flavor QCD~\cite{Takeda:2010cw},
we observe that 
smearing both nucleon source and sink operators is crucial
to identify 
the ground state contribution to $C_{3\rm pt}$.
We employ the Gaussian smearing 
\begin{equation}
 q_{\rm smr}^{\rm gss}({\bf x},t) 
  =
  \sum_{\bfy} \left\{ 
                \left( {\1}+\frac{\omega}{4N}H \right)^N 
               \right\}_{{\bf x,y}} 
  q(\bfy,t), \qquad
  H_{{\bf x,y}} 
  =
  \sum_{i=1}^3 (\delta_{{\bf x,y}-\hat i}+\delta_{{\bf x,y}+\hat i}),
  \label{eqn:Gaussian} 
\end{equation}
where
we omit the gauge links connecting the lattice sites 
$({\bf x},t)$ and $({\bf y},t)$,
which may enhance the statistical fluctuation of $C_{2\rm pt}$ and $C_{3\rm pt}$. 
We use this gauge non-invariant smearing 
on our gauge configurations fixed to the Coulomb gauge.
The parameters $\omega=20$ and $N=400$ are chosen 
by inspecting the plateau of the effective mass of $C_{2\rm pt}$.

\subsection{Setup for the spectrum method}
\label{subsec:meas:spec}

For the spectrum method, 
we use $C_{2\rm pt}$ calculated 
in the course of our study of the light meson spectrum~\cite{Noaki:lat10,m_s}. 
The low-mode contribution $C_{{\rm 2pt}}^{lll}(\bfy,t,\dt)$
is calculated using $N_e\!=\!160$ (80) low-modes 
on the $16^3 \! \times \! 48$ ($24^3 \! \times \! 48$) lattice,
and is averaged over the time-slice $t$
with the spatial source point $\bfy$ kept fixed. 
We use an exponential smearing
\bea
   q_{\rm smr}^{\rm exp}({\bf x},t) 
   =
   \sum_{\bf r} \exp[-0.4|{\bf r}|] q({\bf x}+{\bf r},t) 
\eea
only for the nucleon source operator.
The spatial extent of this smeared operator is roughly equal to 
that of (\ref{eqn:Gaussian}) used for the ratio method.
We observe that the onset of the plateau in the effective mass is
consistent with that of (\ref{eqn:Gaussian}) within the statistical error.

In order to evaluate the derivative $\partial M_N/\partial m_s$ 
in (\ref{eqn:FH}),
we study the $m_s$ dependence of $M_N$ by 
utilizing the reweighting technique~\cite{Hasenfratz:2008fg,DeGrand:2008ps}.
Our Monte Carlo data at the strange quark mass $m_s$ 
are used to estimate the two-point function 
at a slightly shifted strange quark mass $m_s^\prime$ as 
\bea
  \langle C_{2{\rm pt}} \rangle_{m_s^\prime}
  & = & 
  \langle C_{2{\rm pt}} \, \tilde{w}(m_s^\prime,m_s) \rangle_{m_s},
   \label{eq:rew}
\eea
where $\langle \cdots \rangle_{m_s}$ represents the Monte Carlo average
at $m_s$, 
and $\tilde{w}$ is the reweighting factor for a given configuration
\bea
   \tilde{w}(m_s^\prime,m_s)
   & = & 
   \frac{w(m_s^\prime,m_s)}{\langle w(m_s^\prime,m_s) \rangle_{m_s}},
   \hspace{5mm}
   w(m_s^\prime,m_s)
   = 
   \det\left[ \frac{D(m_s^\prime)}{D(m_s)} \right].
   \label{eqn:rw_fctr}
\eea

Similarly to $S^{\rm lat}$ and $C_{2{\rm pt}}$,
$w$ can be decomposed into contributions from low- and high-modes
\bea
   w(m_s^\prime,m_s)
   & = & 
   w_{\rm low}(m_s^\prime,m_s) \,w_{\rm high}(m_s^\prime,m_s),
   \\
   w_{\rm low (high)}(m_s^\prime,m_s) 
   & = & 
   \det\left[ 
      {\mathcal P}_{\rm low (high)} 
      \frac{D(m_s^\prime)}{D(m_s)} 
      {\mathcal P}_{\rm low (high)}
   \right].
\eea
We exactly calculate $w_{\rm low}$ using the low-lying eigenvalues,
whereas $w_{\rm high}$ is estimated 
by a stochastic estimator for its square 
\begin{eqnarray}
 w_{\rm{high}}^2(m_s^\prime,m_s)
  =
  \frac{1}{N_r}
  \sum_{r=1}^{N_r} e^{-\frac{1}{2}({\mathcal P}_{\rm high}\xi_r)^\dagger 
                             (\Omega-1) {\mathcal P}_{\rm high} \xi_r}.
  \label{eq:high}
\end{eqnarray}
Here 
$ \Omega 
  \equiv D(m_s)^\dagger \{D(m_s^\prime)^{-1}\}^\dagger D(m_s^\prime)^{-1}D(m_s)$,
and 
$\{\xi_1,...,\xi_{N_r}\}$ is a set of pseudo-fermion fields 
whose elements are generated with the Gaussian probability.  

\begin{figure}[tbp]
 \centering
 \includegraphics[width=0.48\textwidth,clip]{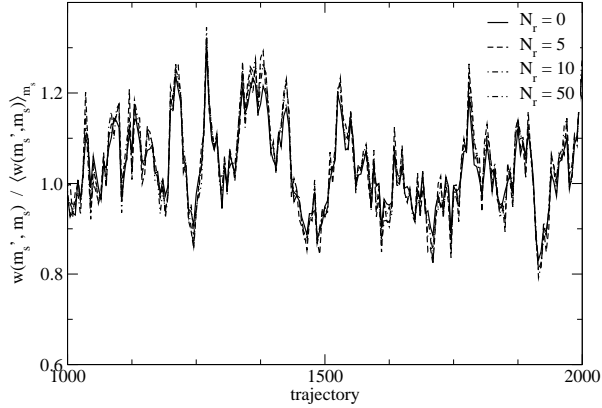}
 \caption{
 Monte Carlo history of reweighting factor $\tilde{w}(m_s^\prime,m_s)$
 to shift the strange quark mass from $m_s^\prime\!=\!0.080$ to $m_s\!=\!0.075$
 at $m_{ud}\!=\!0.050$.
 Different lines show 
 data calculated with different numbers of the pseudo-fermion fields $N_r$.
 }
 \label{fig:rwfctr}
\end{figure}

An important practical issue is 
how many pseudo-fermion fields are needed to reliably estimate $w_{\rm{high}}$.
Since $w_{\rm high}$ is a product of $12N_s^3N_t-N_e$ eigenvalues,  
it largely deviates from unity unless $m_s\!\simeq\!m_s^\prime$.
We observe, however, that it has small statistical fluctuation,
after taking the ratio $\tilde w(m_s',m_s)=w(m_s',m_s)/\langle
w(m_s',m_s)\rangle_{m_s}$.
Consequently,
the normalized reweighting factor $\tilde{w}$
is essentially controlled by the low-mode contribution $w_{\rm low}$.
We therefore do not need large number of the pseudo-fermion fields 
to estimate $w_{\rm high}$ 
as demonstrated in Fig.~\ref{fig:rwfctr}. 

In this study, we reweight $C_{2\rm pt}$ at $m_s\!=\!0.080$ 
to 20 different values 
\bea
   m_s^\prime 
   & = & 
   0.0600,\, 0.0650,\, 0.0700,\, 0.0725,\, 0.0750,\, 0.0775,\,
   0.0780,\, 0.0785,\, 0.0790,\, 0.0795,\,
   \notag \\
   &&
   0.0805,\, 0.0810,\, 0.0815,\, 0.0820,\, 0.0825,\,
   0.0850,\, 0.0875,\, 0.0900,\, 0.0950,\,
   0.1000.
\eea
We shift these values by $+0.020$ when we reweight $C_{2\rm pt}$ at
$m_s\!=\!0.100$.  
These values roughly cover a region 
$m_s^\prime\!\in\![m_s\!-\!25~\mbox{MeV},m_s\!+\!25~\mbox{MeV}]$,
where the low-mode dominance of $\tilde{w}$ is confirmed. 
We set $N_r\!=\!5$ in the whole region of $m_s^\prime$. 
\section{Results at the simulated quark masses}
\label{Sec:result}


In the following subsections,
we present our results for $\NslatN$ 
obtained 
at simulated quark masses 
by using the ratio and spectrum methods.
Note that $\NslatN$ represents the bare value on the lattice, 
and results for the renormalization group invariant parameter 
$f_{T_s}$ will be given in the next section.

\subsection{Ratio method}

\begin{figure}[tbp]
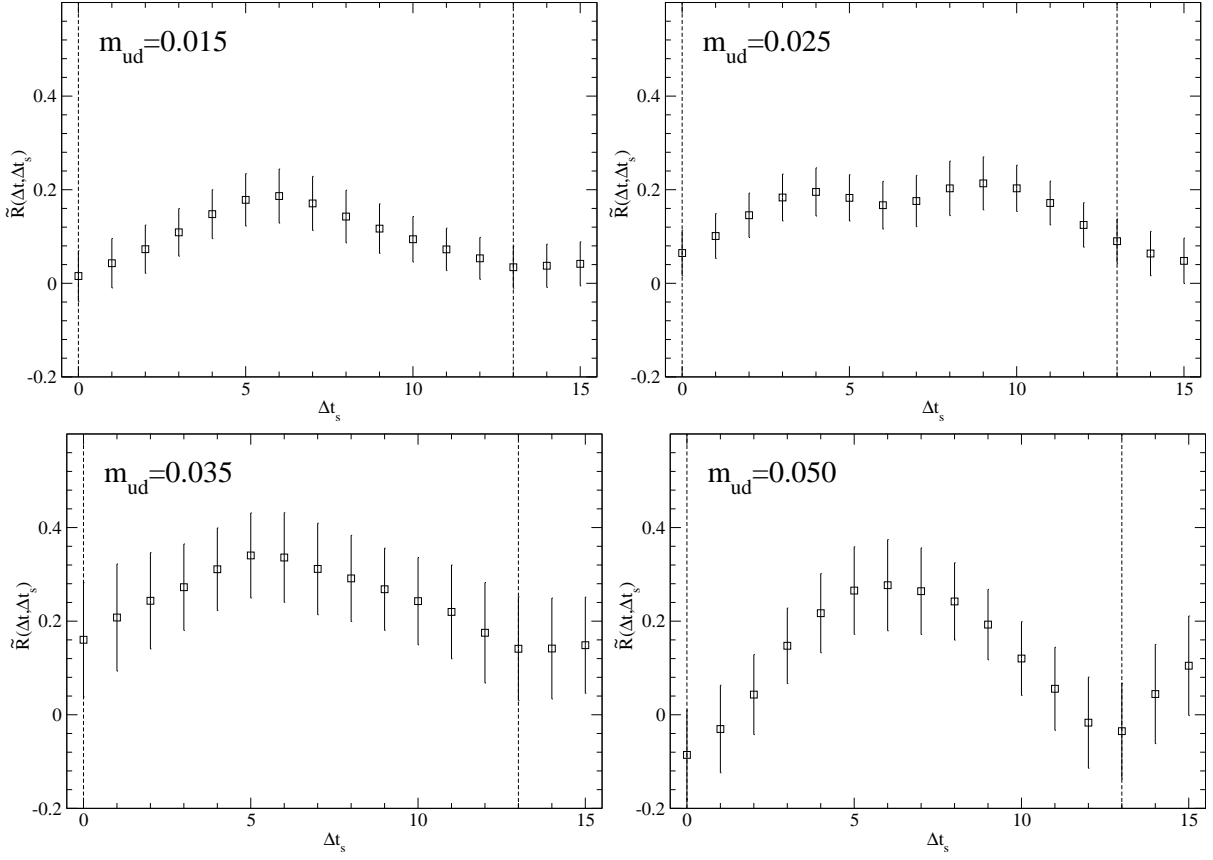

  \centering
  \includegraphics[width=0.48\textwidth,clip]{fig3a.eps}
  \includegraphics[width=0.48\textwidth,clip]{fig3b.eps}
  \vspace{1mm}

  \includegraphics[width=0.48\textwidth,clip]{fig3c.eps}
  \includegraphics[width=0.48\textwidth,clip]{fig3d.eps}
 \caption{
 Approximated ratio $\tilde{R}(\dt,\dts)$ at $m_s\!=\!0.080$
 as a function of $\dts$.
 We plot results obtained at different values of $m_{ud}$ in the four panels. 
 The vertical dashed lines show the locations of the nucleon source 
 and sink operators. 
 }
  \label{Fig:ratio_l}
\end{figure}

We extract $\NslatN$ from the ratio of $C_{\rm 3pt}(\dt,\dts)$ and $C_{\rm
2pt}(\dt)$ 
\begin{equation}
    R(\dt,\dts) 
   \equiv \frac{C_{\rm 3pt}(\dt,\dts)}{C_{\rm 2pt}(\dt)}
   \xrightarrow[\dt, \dts \to \infty\\ ]{ } 
   \NslatN,
   \label{eqn:ratio}
\end{equation}
where 
$\dt$ is the temporal interval between the nucleon source and sink. 
The scalar quark loop $S^{\rm lat}$ is set on the time-slice apart
from the nucleon source by $\dts$. 
Note that 
$C_{\rm 3pt}(\dt,\dts)$ and $C_{\rm 2pt}(\dt)$ are calculated using LMA
and, hence,
we suppress the coordinates of the nucleon source,
namely $(\bfy,t)$ in (\ref{eqn:2pt}) and (\ref{eqn:3pt_pieces}).


The ratio $R$ may receive contamination from excited states of the nucleon
when the temporal separation $\dt$ is not sufficiently large and/or 
the scalar operator is too close to the nucleon operators 
($\dts \sim 0$ or $\dt$).
We therefore need to identify a plateau of $R(\dt,\dts)$,
where the excited state contamination is negligible.
To this end,
we consider the same ratio 
but approximated by taking only the low-mode contribution 
$S_{\rm low}^{\rm lat}$ for the quark loop $S^{\rm lat}$
in (\ref{eqn:3pt_loop_LH}).
This approximated ratio, which we denote by $\tilde{R}$ in the following, 
is useful to identify the plateau of $R$,
because 
i) $R$ is well dominated by the low-mode approximation $\tilde{R}$,  
and 
ii) $\tilde{R}$ is free from a large noise due to the stochastic method
to estimate $S_{\rm high}^{\rm lat}$, 
which obscures the excited state contamination.
We refer the reader to Ref.~\cite{Takeda:2010cw}
for a more detailed discussion.


\begin{figure}[tbp]
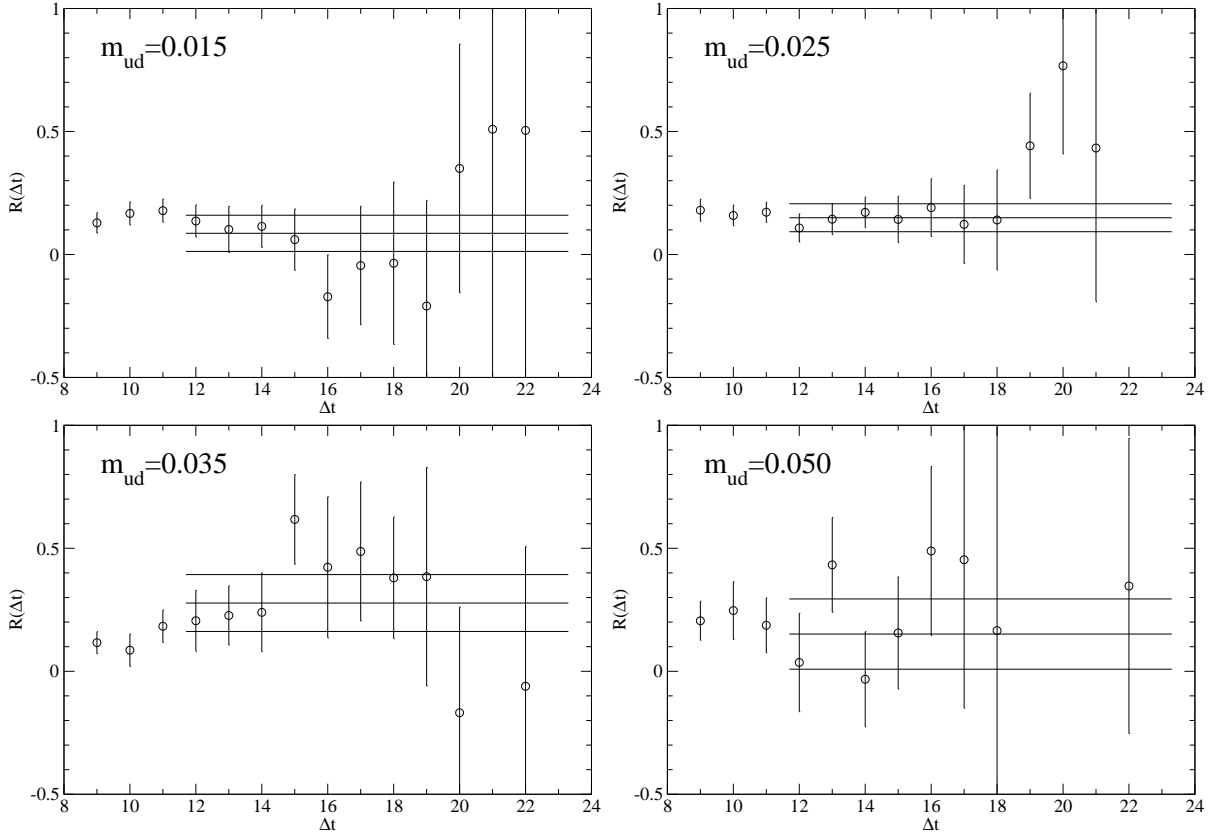

 \centering
 \includegraphics[width=0.48\textwidth,clip]{fig4a.eps}
 \includegraphics[width=0.48\textwidth,clip]{fig4b.eps}
 \vspace{3mm}
 \includegraphics[width=0.48\textwidth,clip]{fig4c.eps}
 \includegraphics[width=0.48\textwidth,clip]{fig4d.eps}
 \caption{
 Results of the constant fit to $R(\dt,\dts)$ as a function of $\dt$. 
 Four panels show results obtained at different values of $m_{ud}$
 and at $m_s\!=\!0.080$. 
 }
 \label{Fig:dtfit}
\end{figure}

Figure~\ref{Fig:ratio_l} shows 
$\tilde{R}(\dt,\dts)$ with a fixed value of $\dt\!=\!13$ 
as a function of $\dts$.
We obtain nonzero signal for $\tilde{R}(\dt,\dts)$, 
which do not show significant $\dts$ dependence 
at $\dts \!\sim\! \dt/2$. It implies that the scalar operator is
sufficiently far from nucleon operators. 

We then carry out a constant fit to 
the ratio without the approximation $R(\dt,\dts)$
using a fit range of $\dts=[5,\dt\!-\!5]$ for each $\dt$. 
As plotted in Fig.~\ref{Fig:dtfit}, the fit results do not show 
statistically significant dependence on $\Delta t$ at $\Delta t \!\geq\! 12$, 
that indicates that the data are dominated by the ground state
contribution. Although the statistical signal is worse at
$m_{ud}\!=\!0.050$, it is reasonable to assume the ground state 
saturation at around the same $\Delta t$ region as other $m_{ud}$'s.

From these observations on the $\dts$ and $\dt$ dependences, 
we determine $\NslatN$ by a simultaneous constant fit to $R(\dt,\dts)$
with fit ranges of $\dts=[5,\dt\!-\!5]$ and $\dt=[12,23]$.
The numerical results are listed in Table~\ref{tab:ratio}. 
We also test a fitting form taking account of the first excited state 
with a slightly wider fit range of $\dts$.
This fit yields $\NslatN$ in good agreement with those from the constant fit,
because the excited state contribution is small 
as expected from the mild $\dt$ dependence of $R(\dt,\dts)$.

We repeat the same analysis at two smallest $m_{ud}$'s 
but on the smaller volume $16^3 \! \times \! 48$. 
The numerical results are listed in Table~\ref{tab:ratio_FVEs}. 
The difference in $\NslatN$ between the two volumes are 
well below our statistical accuracy 
suggesting that finite volume effects (FVEs)
can be neglected within the statistical error.
We therefore use the numerical results in Table~\ref{tab:ratio}
in the chiral extrapolation to determine $\NslatN$ at physical quark masses.

 \begin{table}[tbp]
  \begin{center}
   \begin{tabular}{c|l|l|ll|ll} \hline\hline
    $m_{ud}$   & 0.015   & 0.025   & 0.035   & 0.035 & 0.050 & 0.050 \\
    $m_{s}$    & 0.080   & 0.080   & 0.080   & 0.100 & 0.080 & 0.100 \\\hline
    $\NslatN$ & 0.09(7) & 0.15(6) & 0.28(12) &0.14(10) &0.15(14) &0.20(18) \\
    \hline\hline
   \end{tabular}
  \end{center}
  \caption{
    Strange quark content $\NslatN$ calculated in the ratio method.
  } 
  \label{tab:ratio}
 \end{table}

 \begin{table}[tbp]
  \begin{center}
   \begin{tabular}{c|ll|ll} \hline\hline
    $m_{ud}$ & 0.015 & 0.015 &0.025 & 0.025 \\ 
    $m_{s}$  & 0.080 &0.100 &0.080 &0.100 \\\hline
    $\NslatN$ & 0.34(24) &0.29(32) &0.21(16) &0.02(9) \\
    \hline\hline
   \end{tabular}
  \end{center}
  \caption{
    Same as Table~\ref{tab:ratio} 
    but for $m_{ud}\!=\!0.015$ and $0.025$ 
    on the smaller volume $16^3 \! \times \! 48$.
  } 
  \label{tab:ratio_FVEs}
 \end{table}

\subsection{Spectrum method}


\begin{figure}[tbp]
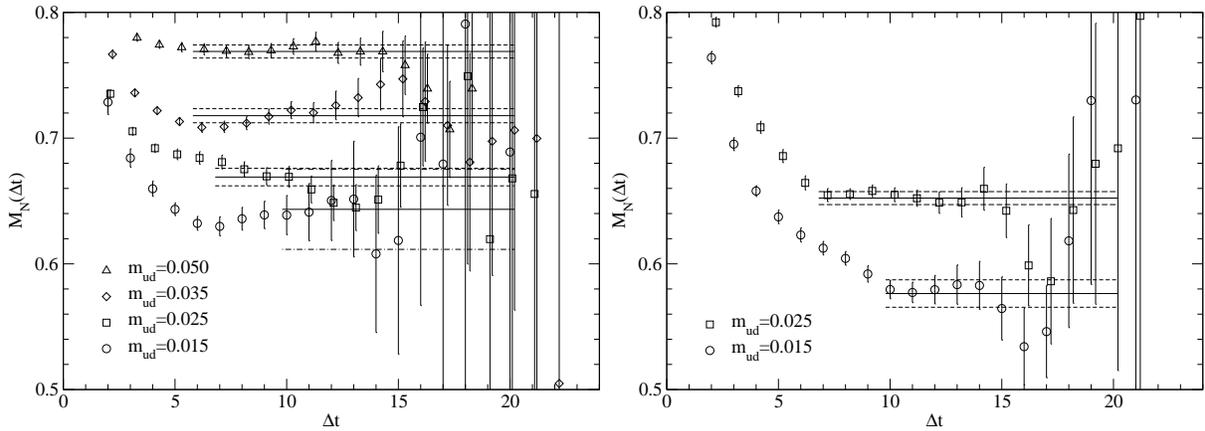

  \centering
    \includegraphics[width=0.48\textwidth,clip]{fig5a.eps}
    \includegraphics[width=0.48\textwidth,clip]{fig5b.eps}
\caption{
 Nucleon effective masses at $m_s\!=\!0.080$. 
 Left and right panels show 
 results on the $16^3 \times 48$ and $24^3 \times 48$ lattices,
 respectively.
 Horizontal lines show $M_N$ obtained from a single exponential fit 
 to $C_{2\rm pt}(\dt)$.
}
\label{fig:eff_mass}
\end{figure}

\begin{figure}[t]
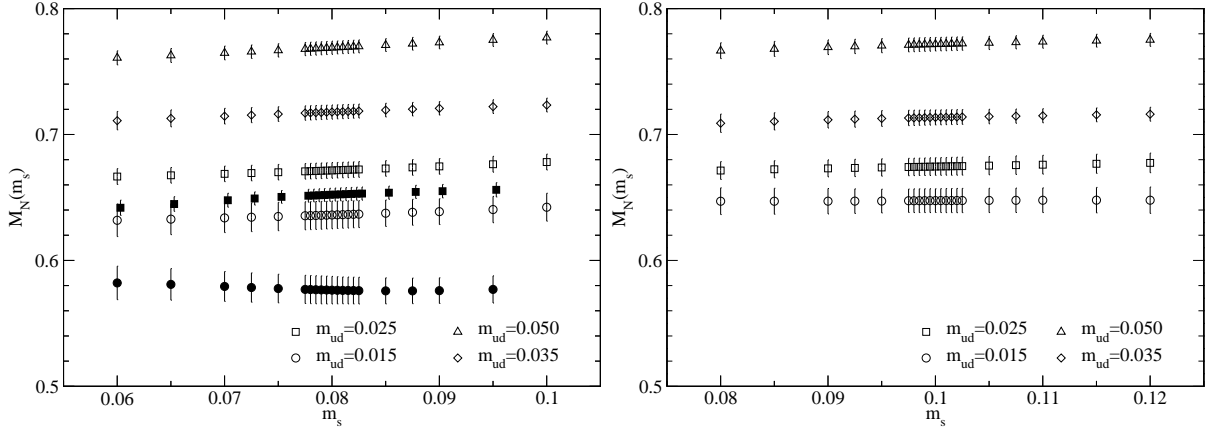

 \centering
 \includegraphics[width=0.48\textwidth,clip]{fig6a.eps}
 \includegraphics[width=0.48\textwidth,clip]{fig6b.eps}
 \caption{
 Nucleon masses $M_N$ as a function of $m_s$. 
 Left and right panels show $M_N$ obtained by reweighting from 
 that at $m_s\!=\!0.080$ and 0.100, respectively.
 We plot results on the $16^3 \times 48$ and $24^3 \times 48$ lattices, 
 by open and filled symbols.
 Filled squares are slightly shifted in the horizontal direction for clarity. 
 }
 \label{fig:slope}
\end{figure}

In the spectrum method, 
we evaluate $\NslatN$ from the $m_s$ dependence of the nucleon mass $M_N$. 
Figure~\ref{fig:eff_mass} shows 
examples of the nucleon effective mass obtained at $m_s\!=\!0.080$.
By a single exponential fit $C_{2\rm pt}(\dt) \! \propto \! e^{-M_N \dt}$, 
we determine $M_N$ with an accuracy of 2\% (0.8\%) 
at our smallest (largest) $m_{ud}$ on $24^3 \times 48$ ($16^3 \times 48$).


The FVE in $M_N$ at $m_{ud}\!=\!0.025$ is not statistically significant: 
it is only 2\,$\sigma$ (3\,\%) effect. 
We expect similarly small effect at heavier $m_{ud}$'s.
The magnitude of the FVE at $m_{ud}\!=\!0.015$ is difficult to estimate 
due to a large statistical error of $M_N$ 
on the smaller volume $16^3 \times 48$.
We note that 
the FVE at $m_{ud}=0.015$ on the larger volume $24^3 \times 48$
is estimated as 0.7\,\% 
from $SU(2)$ heavy baryon chiral perturbation theory (HBChPT) at one-loop.
In addition, 
it is plausible that the FVE has a mild dependence on $m_s$
leading to small effect in $\NslatN$.


\begin{figure}[t]
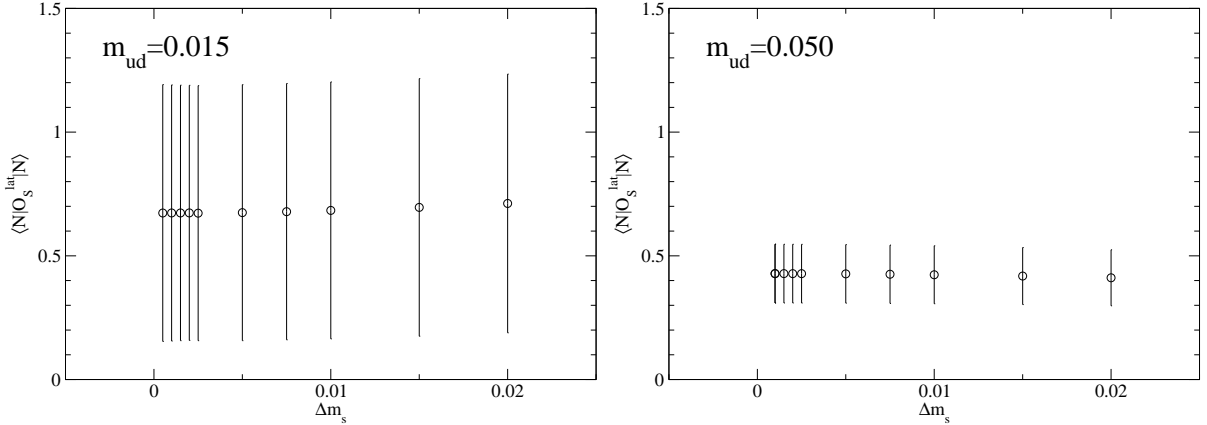

  \centering
  \includegraphics[width=0.48\textwidth,clip]{fig7a.eps}
  \includegraphics[width=0.48\textwidth,clip]{fig7b.eps}
  \caption{
    Fitted results for $\NslatN$ 
    as a function of the width of the fitting range $\Delta m_s$. 
    Left and right panels show results at 
    $(m_{ud},m_s)\!=\!(0.015,0.080)$ and $(m_{ud},m_s)\!=\!(0.050,0.080)$,
    respectively.
  }
  \label{Fig:NslatN_vs_dms}
\end{figure}

\begin{table}[t]
 \begin{center}
 \begin{tabular}{r|l|l|ll|ll} \hline\hline
    $m_{ud}$   & 0.015 & 0.025 & 0.035 & 0.035 & 0.050 & 0.050 
              \\
    $m_{s}$    & 0.080 & 0.080 & 0.080 & 0.100 & 0.080 & 0.100 
              \\ \hline
    $\NslatN$ & $-$0.16(35) & 0.35(13) &0.31(15) &0.16(12) &0.42(10) &0.22(10) 
              \\
    \hline\hline
 \end{tabular}
 \end{center}
 \caption{
    Strange quark content $\NslatN$ obtained from the spectrum method.
 } 
 \label{tab:strange}
\end{table}

\begin{table}[t]
 \begin{center}
 \begin{tabular}{r|ll|ll} \hline\hline
    $m_{ud}$ & 0.015 & 0.015 & 0.025 & 0.025 \\ 
    $m_{s}$  & 0.080 &0.100 &0.080 &0.100 \\\hline
    $\NslatN$ & 0.68(52) &$-$0.60(55) &0.34(11) &0.12(18) \\
    \hline\hline
 \end{tabular}
 \end{center}
 \caption{
    Same as Table~\ref{tab:strange} 
    but at $m_{ud}\!=\!0.015$ and 0.025 
    on the smaller volume $16^3 \! \times \! 48$.
 } 
 \label{tab:strange_FVEs}
\end{table}

As explained in the previous section, 
we calculate $M_N$ at shifted values of $m_s$
by exploiting the reweighting technique. 
Results are plotted as a function of $m_s$ in Fig.~\ref{fig:slope}.
We successfully reweight our data to $m_s \pm 0.02$ ($\pm 25$~MeV).
Namely, the reweighting does not largely increase 
the statistical error of $M_N$.
This is because 
i) the reweighting factor $\tilde{w}$, is accurately calculated 
with the small number of the noise samples, 
as discussed in the previous section,
and ii) resulting values are typically $O(1)$ as plotted in
Fig.~\ref{fig:rwfctr}.


We extract the slope $\rd M_N/\rd m_s$ by fitting $M_N$
in the region of $[m_s-\Delta m_s,m_s+\Delta m_s]$ with $\Delta m_s\!=\!0.010$
to a linear form 
\bea
  M_N = d + \NslatN m_s.
  \label{eqn:spec_fit}
\eea
The numerical results are summarized in Table.~\ref{tab:strange}. 
Figure~\ref{Fig:NslatN_vs_dms}
shows that the fitted result for $\NslatN$ 
is stable against the choice of the fitting range $\Delta m_s$
as expected from the mild $m_s$ dependence of $M_N$ 
shown in Fig.~\ref{fig:slope}. 
We also confirm that 
adding higher order terms to (\ref{eqn:spec_fit}) 
does not change $\NslatN$ significantly.

In order to directly check FVEs to $\NslatN$,
we repeat the analysis at two lightest $m_{ud}$ but on the smaller volume.
A comparison with results listed in Table~\ref{tab:strange_FVEs}
suggests that FVE is not significant with our statistical accuracy,
which is consistent with our observation in the ratio method.


\subsection{Comparison between two methods}
\label{subsec:ratio:comp}

\begin{figure}[t]
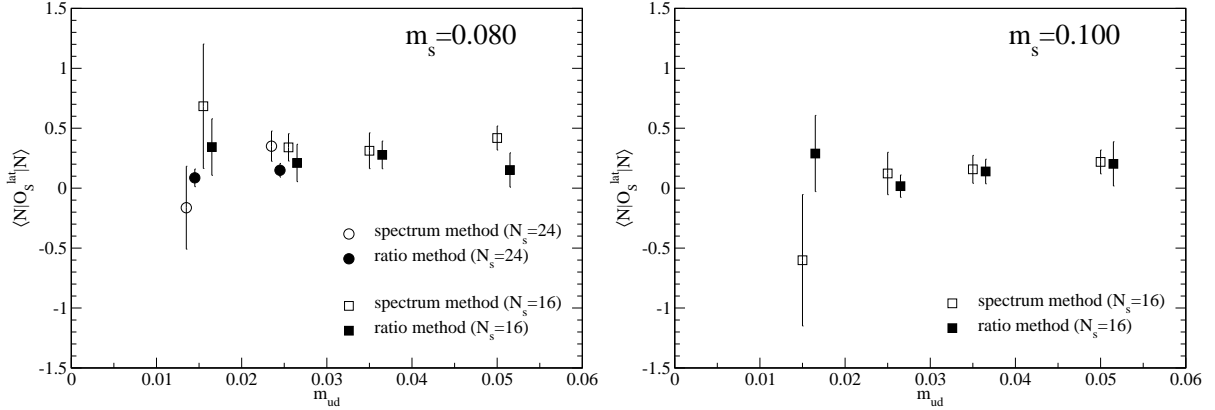

  \centering
  \includegraphics[width=0.48\textwidth,clip]{fig8a.eps}
  \includegraphics[width=0.48\textwidth,clip]{fig8b.eps}
  \caption{
    Strange quark content $\NslatN$ as a function
    of $m_{ud}$ (given in the lattice unit).
    Left and right panels show the results from the spectrum and ratio
    methods, respectively.
  }
  \label{fig:mass}
\end{figure}


Figure~\ref{fig:mass}
compares $\NslatN$ obtained from the spectrum and ratio methods.
We observe a good agreement between the two methods.
The same figure also shows that
FVEs in $\NslatN$ are not significant at the two smallest $m_{ud}$'s
as already mentioned in the previous subsections. 
These observations suggest that 
systematics of our determinations at given quark masses $(m_{ud},m_s)$ 
is not substantial.


With our simulation setup,
the accuracy at two heaviest $m_{ud}$'s are comparable between the two
methods, while
the ratio method provides a more accurate determination 
at lighter $m_{ud}$'s.
This is mainly because i) we use the better setup of LMA for the ratio method 
and ii) the volume size is increased at these $m_{ud}$'s.
For instance, 
we average $C_{{\rm 2pt}}^{lll}(\bfy,t,\dt)$ at the 16 choices of 
the spatial location $\bfy$,
while $\bfy$ is kept fixed in the spectrum method.
Our data at the 16 choices listed in (\ref{label:lma:spatial:ratio})
have less correlation among them on a larger volume 
and hence LMA works better.

\section{Chiral extrapolation}
\label{Sec:Chiral}

\begin{figure}[tbp]
  \centering
  \includegraphics[width=0.48\textwidth,clip]{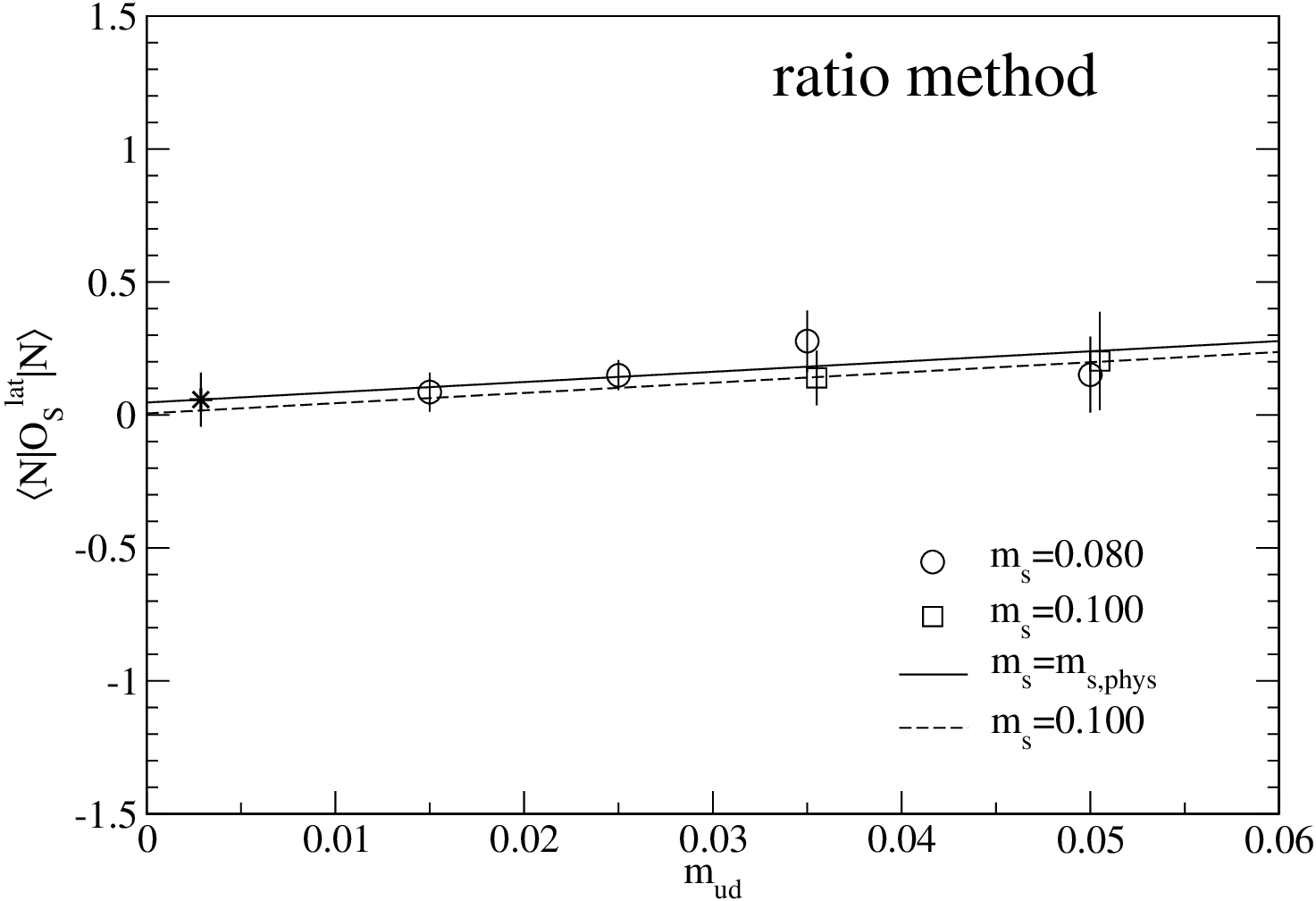}
  \includegraphics[width=0.48\textwidth,clip]{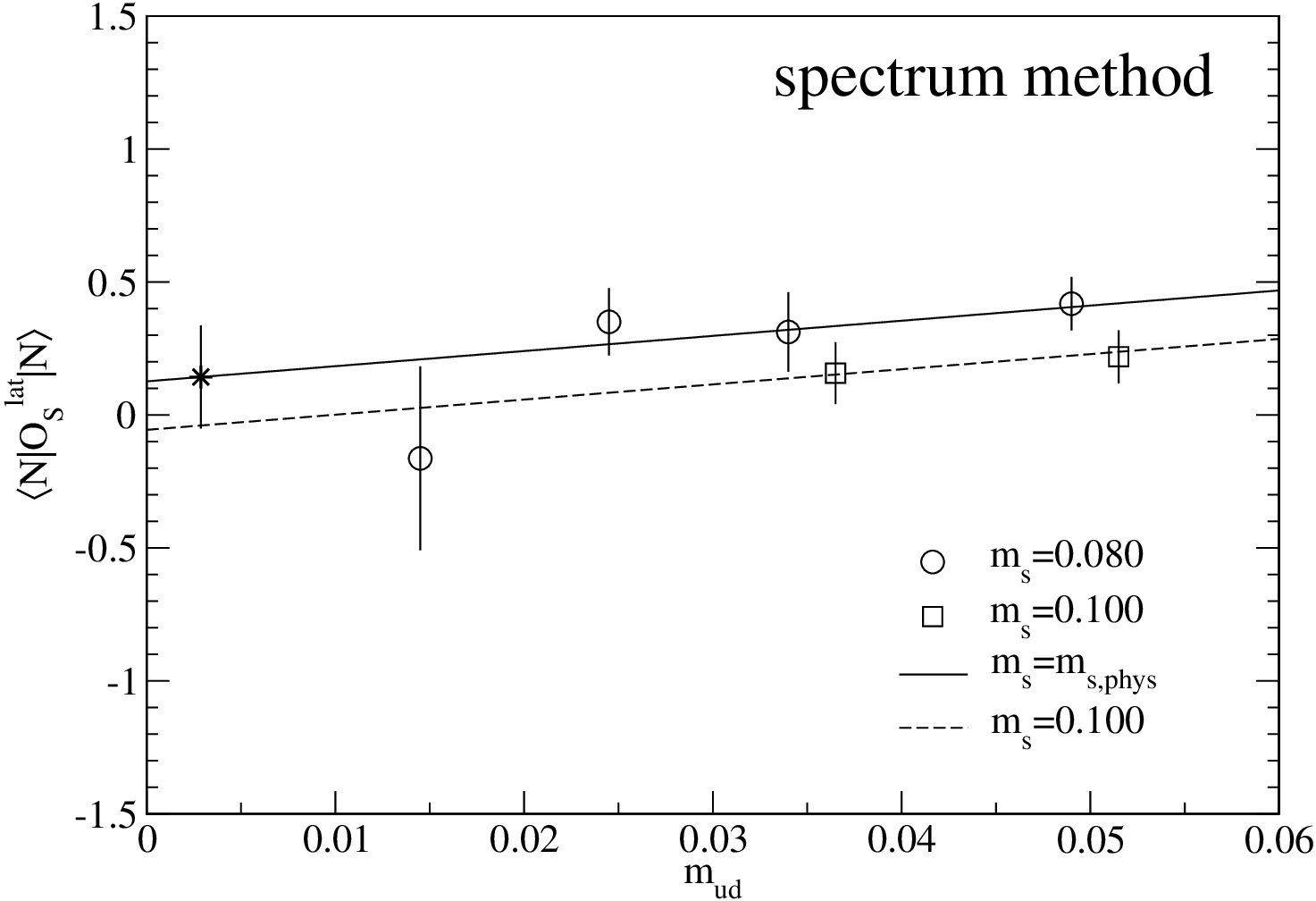}
  \caption{
    Linear extrapolations of $\NslatN$ obtained from 
    ratio (left panel) and spectrum (right panel) methods. 
    Solid and dashed lines show the fit lines 
    at $m_s\!=\!m_{s,\rm phys}$ and 0.100, respectively.
    We omit the fit line at $m_s\!=\!0.080$, which is not 
    indistinguishable from that at $m_s\!=\!m_{s,\rm phys}$
    in the scale of the figure. 
    Star symbols represent $\NslatN$ extrapolated to the physical quark masses.
    Square symbols are slightly shifted in the horizontal direction
    for clarity.
  }
  \label{Fig:linear}
\end{figure}

\begin{table}[tbp]
  \centering
  \begin{tabular}{lcccccc}
    \hline\hline
    & $\chi^2/{\rm d.o.f.}$  & $\rm d.o.f.$  & $c_0$  & $c_{1}$ 
    & $c_{1,s}$ & $\NslatN$ \\ 
    \hline
    spectrum method & 0.54 & 3   & 0.90(47)   & 5.7(5.1)  
                                 & $-$9.6(5.4)  & 0.15(0.19) \\
    \hline    
    ratio method    & 0.38 & 3   & 0.22(43)   & 3.9(4.1) 
                                 & $-$2.2(5.7)  & 0.058(0.101) \\ 
    \hline\hline
  \end{tabular}
  \caption{
    Numerical results of linear chiral extrapolation. 
  } 
  \label{tab:linear}
\end{table}

In Fig.~\ref{Fig:linear}, 
we plot $\NslatN$ obtained from the two methods as a function of $m_{ud}$. 
Note that our data cover a region of $M_\pi \!\sim $~300\,--\,540~MeV,
and our lighter strange quark mass $m_s\!=\!0.080$ 
is already close to the physical mass $m_{s,\rm phys}=0.081$. 
The figure shows that $\NslatN$ has a very mild dependence
on both $m_{ud}$ and $m_s$, which has also been observed 
in our previous study in two-flavor QCD~\cite{Takeda:2010cw}. 
Our data are well described by a linear fit
\bea
  \NslatN = c_0 + c_{1,ud} m_{ud} + c_{1,s} m_{s}
  \label{eqn:linear}
\eea
as plotted in the same figure. 
Numerical results of the fit are summarized in Table~\ref{tab:linear}.
We also confirm that 
$\NslatN$ at the physical quark masses does not change significantly
by excluding the data at the largest $m_{ud}$ from the fit
and/or by including higher order terms in (\ref{eqn:linear}).

We also test a fitting form based on $SU(3)$ HBChPT
to parametrize the observed quark mass dependence of $\NssN$.
One-loop chiral expansion of $M_N$~\cite{WalkerLoud:2004hf}
and the Feynman-Hellmann theorem~(\ref{eqn:FH}) 
give an expression of $\NslatN$
\begin{equation}
  \NslatN
  = -c_s -B
  \left\{ \frac{3}{2} C_{NNK} \, M_K + 2 C_{NN\eta}\, M_\eta \right\} 
  + c_{2,K} M_K^2 +  c_{2,\eta} M_\eta^2,
  \label{eqn:NNLO}
\end{equation}
where contributions of the decuplet baryons are ignored. 
In this analysis, 
we approximate the higher order corrections 
by the $O(M_{\{K,\eta\}}^2)$ analytic terms. 
Within this approximation, 
we can use the leading order expressions
$M_K^2=B(m_{ud}+m_s)$ and $M_\eta^2=2B(m_{ud}+2m_s)/3$ for the meson masses.
The coefficients $C_{NNK}$ and $C_{NN\eta}$ of the $O(M_{\{K,\eta\}})$ terms 
are written as 
\begin{eqnarray}
  C_{NNK} = \frac{1}{8\pi f^2}\frac{(5D^2-6DF+9F^2)}{3}, \quad 
  C_{NN\eta} = \frac{1}{8\pi f^2}\frac{(D-3F)^2}{6}.
  \label{eqn:NLO:coeff2}
\end{eqnarray}
The axial couplings are fixed to a phenomenological estimate 
$D=0.81$ and $F=0.47$~\cite{Jenkins:1991es} in this analysis.
The low-energy constants in mesonic ChPT, $f$ and $B$,
are set to our lattice estimate 
determined from the meson spectrum and decay constants~\cite{m_s}.

\begin{figure}[tbp]
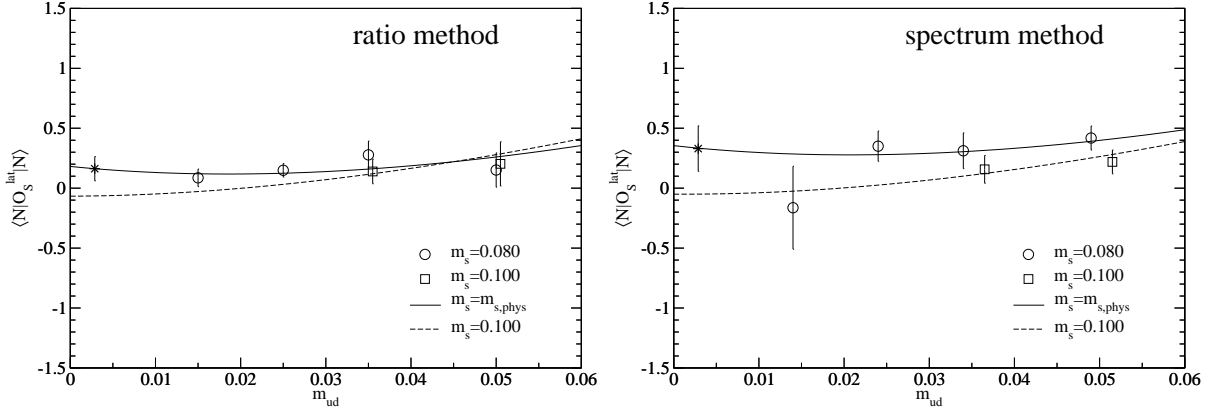

  \centering
  \includegraphics[width=0.48\textwidth,clip]{fig10a.eps}
  \includegraphics[width=0.48\textwidth,clip]{fig10b.eps}
  \caption{
    Chiral fits of $\NslatN$ using $SU(3)$ HBChPT~(\ref{eqn:NNLO}). 
    Left and right panels show fits to $\NslatN$ 
    obtained from the ratio and spectrum methods, respectively.
  }
  \label{Fig:chpt}
\end{figure}

\begin{table}[tbp]
  \begin{center}
  \begin{tabular}{lcccccc} 
    \hline\hline
    & $\chi^2/{\rm d.o.f.}$  & {\rm d.o.f.} 
    & $-c_s$  & $c_{2,K}$ [${\rm GeV}^{-2}$]  & $c_{2,\eta}$ [${\rm GeV}^{-2}$] 
    & $\NslatN$
    \\ \hline
    spectrum method &
    1.2 & 3 & 8.7(5) & 23(4) &$-$6.1(3.6) & 0.33(19)
    \\ \hline
    ratio method & 
    0.75 & 3 & 7.9(4) & 21(3)  & $-$2.8(3.6)  &  0.16(10)
    \\ \hline\hline
    \end{tabular}
  \end{center}
  \caption{ 
    Numerical results of chiral fit using $SU(3)$ HBChPT~(\ref{eqn:NNLO}).
  }
  \label{Tab:chpt}
\end{table}

\begin{figure}[tbp]
  \centering
  \includegraphics[width=0.48\textwidth,clip]{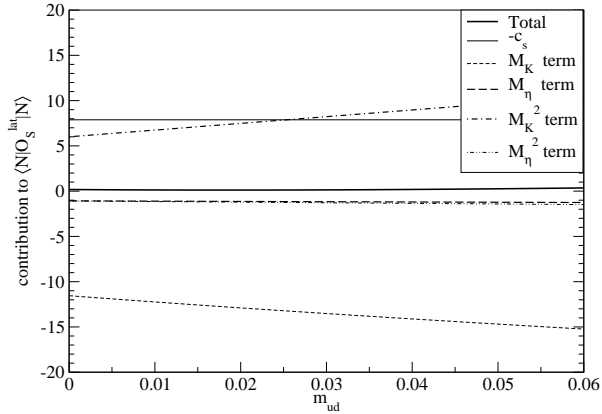}
  \caption{
    Contribution to $\NslatN$ in the chiral expansion~(\ref{eqn:NNLO}). Each
    contribution is calculated at the physical strange quark mass.
  }
  \label{Fig:cntrb}
\end{figure}

The fit using (\ref{eqn:NNLO}) is shown in Fig.~\ref{Fig:chpt}.
Since $\NssN$ depends mildly on $m_{ud}$ through 
the strange meson masses $M_{\{K,\eta\}}$
up to one-loop order of HBChPT,
the mild $m_{ud}$ dependence of our data can be fitted to (\ref{eqn:NNLO})
reasonably well. 
However, numerical results summarized in Table~\ref{Tab:chpt}
suggest a large difference of $\NslatN$ in the $SU(3)$ chiral limit 
between the linear and HBChPT fits 
({\it cf.} $-c_s$ in Table~\ref{Tab:chpt} and $c_0$ in Table~\ref{tab:linear}). 
This is because 
(\ref{eqn:NNLO}) predicts a large $O(M_K)$ contribution to $\NslatN$ 
at $m_{s,\rm phys}$
with the phenomenological estimate of $D$ and $F$.
Then, the fit reproduces our small values of $\NslatN$ 
by a large cancellation among chiral corrections at different orders.
Consequently, 
the HBChPT expansion exhibits a poor convergence 
as shown in Fig.~\ref{Fig:cntrb}.
A similarly poor convergence of HBChPT has been observed 
in our study in two-flavor QCD~\cite{Takeda:2010cw}. 
These observations suggest that, at least for $\NslatN$, 
the $SU(3)$ chiral expansion up to $O(M_{\{K,\eta\}}^2)$
could be applicable only to lattice data at much smaller values of $m_s$.

In this study, therefore, 
we determine $\NslatN$ from the linear fit (\ref{eqn:linear}) 
and use the HBChPT fit only to estimate the systematic uncertainty 
of the chiral extrapolation.
We obtain $\NslatN\!=\!0.15$(19)(18) from
the spectrum method and 0.06(10)(10) from the ratio method.
The first and second errors represent the statistical and systematic ones.
In this study,
the ratio method provides a statistically better determination of $\NssN$.
This is partly because we employ a better setup of LMA for the ratio method
as mentioned in subsections~\ref{subsec:meas:ratio} and \ref{subsec:ratio:comp}.
A nucleon operator with a better overlap with the nucleon ground state 
also improves the accuracy of the spectrum method.

As discussed in Sec.~\ref{Sec:result}, 
we expect that the FVE on our larger volume is small.
The discretization effect is estimated as 
$O((a\Lambda)^2)\sim 9\%$ from a simple order counting 
using $\Lambda=500$ MeV. 
These systematic errors are well below our statistical accuracy 
and, hence, ignored in the following discussions.
We also note that 
exact chiral symmetry in our simulation,
forbids the mixing with the light quark contents 
$\NllN$~\cite{Michael:2001bv,Ohki:2008ff,Takeda:2010cw},
which turned out to introduce a large uncertainty 
in $\NssN$~\cite{Michael:2001bv}.

\begin{figure}[tbp]
  \centering
  \includegraphics[width=0.8\textwidth,clip]{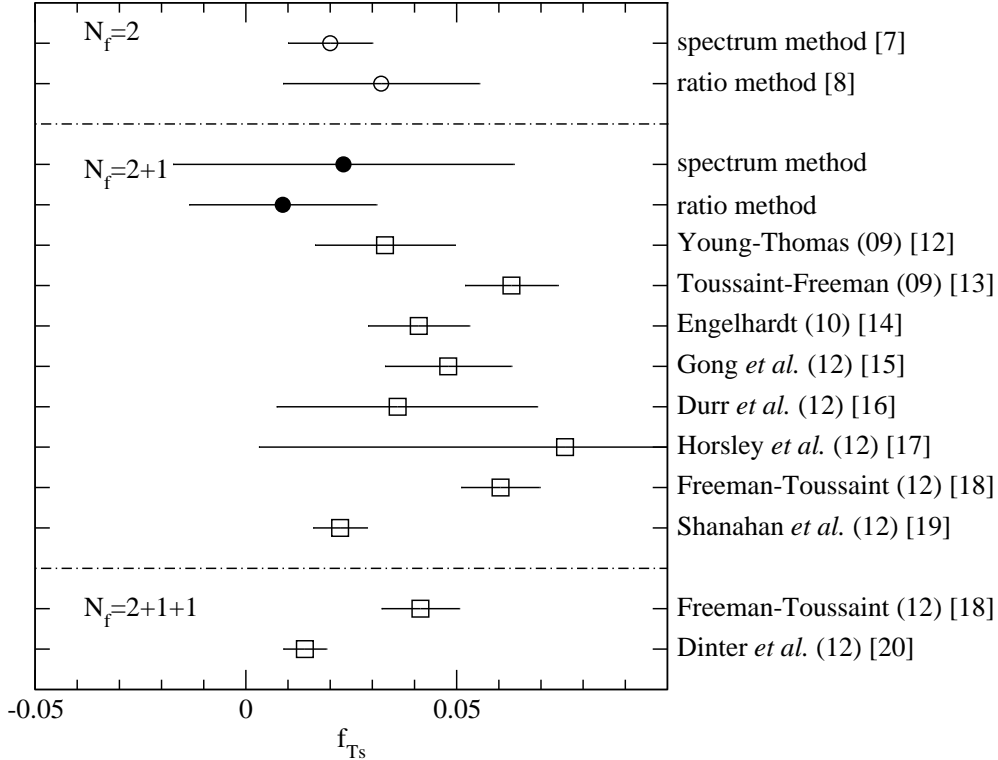}
  \caption{
    Comparison of $f_{T_s}$ with our previous studies in $N_f\!=\!2$ QCD
    and recent studies in $N_f\!=\!2+1$ and 2+1+1
 QCD~\cite{Young:2009zb,Toussaint:2009pz,Engelhardt:2010zr,Durr:2011mp,Horsley:2011wr,Dinter:2012tt,Freeman:2012ry,Gong:2012nw,Shanahan:2012}. 
 We convert $m_s\NssN$ or $\NssN$
 in~\cite{Toussaint:2009pz,Horsley:2011wr,Freeman:2012ry,Shanahan:2012} to $f_{T_s}$ using the  experimental value of $M_N$ and
 $m_s$ obtained in~\cite{Bazavov:2009tw}. 
  }
  \label{Fig:comparison}
\end{figure}

\begin{figure}[tbp]
  \centering
  \includegraphics[width=0.6\textwidth,clip]{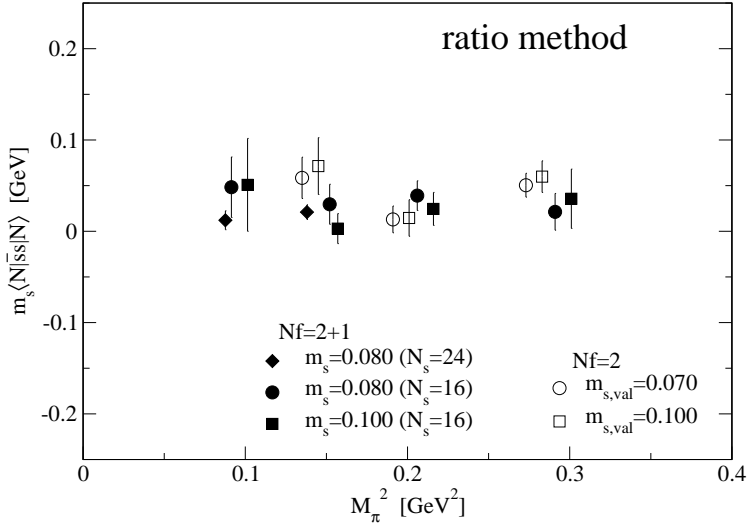}
  \caption{
    A comparison of a renormalization invariant quantity $m_s \NssN$ 
    between $N_f\!=\!2$ (Refs.~\cite{Takeda:2010cw}, open symbols)
    and 2+1 (this study, filled symbols) QCD. 
    We plot data from the ratio method as a function of $M_\pi^2$.
  }
  \label{Fig:comparison_methods}
\end{figure}

The bare matrix element $\NslatN$ is converted to 
the renormalization invariant parameter
\begin{eqnarray}
  f_{T_s}
  \equiv
  \frac{m_s \NslatN}{M_N} =
  \begin{cases}
   0.023(29)(28) & \quad \text{(spectrum method)}, \\
   0.009(15)(16) & \quad \text{(ratio method)}, 
  \end{cases}
\end{eqnarray}
where we use the experimental value of $M_N$. 
In Fig.~\ref{Fig:comparison}, we compare our results of $f_{T_s}$ 
with our previous estimate in $N_f\!=\!2$ QCD~\cite{Ohki:2008ff,Takeda:2010cw}.
All of our studies give consistent results for $f_{T_s}$. 
As confirmed in Fig.~\ref{Fig:comparison_methods}, 
sea strange quark loops have small effects 
to a renormalization invariant quantity $m_s \NslatN$
leading to the good agreement between $N_f\!=\!2$ and 2+1 QCD. 
As mentioned in the Introduction, 
our previous study using the spectrum method 
in $N_f\!=\!2$ QCD~\cite{Ohki:2008ff}
estimated $\NssN$ from the derivative $\partial M_N/\partial m_{ud,\rm sea}$ 
with $m_{ud,\rm \{sea,val\}}$ sending to $m_{s,\rm phys}$.
This turns out to be a reasonable estimate of $\NssN$
because of the very mild dependence of $\NssN$ on $m_{\{ud,s\}}$ 
shown in Fig.~\ref{Fig:linear} 
as well as the small effect of dynamical strange quark loops 
in Fig.~\ref{Fig:comparison_methods}.

Figure~\ref{Fig:comparison} also compares our results with 
recent studies 
in $N_f\!=\!2+1$ and 2+1+1 lattice
QCD~\cite{Young:2009zb,Toussaint:2009pz,Engelhardt:2010zr,Durr:2011mp,Horsley:2011wr,Dinter:2012tt,Freeman:2012ry,Gong:2012nw}. 
All these studies favor small strange quark content $f_{T_s}\! \lesssim 0.1$. 
Strictly speaking, 
the results of Ref.~\cite{Toussaint:2009pz}
appears to be slightly higher (2.5\,$\sigma$) than our best estimate,
that is $f_{T_s}$ in $N_f\!=\!2+1$ QCD from the ratio method. Recently,
the same authors present improved estimates in $N_f=2+1$ and
$2+1+1$ QCD~\cite{Freeman:2012ry}. These results also indicate a
slightly large value of $f_{T_s}\!\sim 0.06$.
Given the large statistical errors, however, the difference is not very
significant. 

Compared to these lattice estimates,
the phenomenological studies \cite{piN:exprt,sigma_0:hbchpt}
predict a rather large estimate $0.41(9)$ based on HBChPT up to
quadratic order in the quark masses.
The poor convergence of our chiral fit based on the same effective theory
suggests that its convergence at physical quark masses 
should be carefully examined.

\section{Conclusion}
\label{Sec:Conclusion}

In this paper, we calculate the strange quark content of the nucleon
in $2+1$\,-flavor lattice QCD. 
Two determinations using the ratio and spectrum methods
as well as our previous studies in two-flavor QCD
consistently favor a small strange quark content $f_{T_s}\!\lesssim\!0.05$.
In contrast, 
phenomenological studies based on HBChPT 
have led to a rather larger value $0.1$\,--\,0.7
which is, however, unexpectedly large 
as a content of sea quarks of a single flavor.


In this study,
we utilize several simulation techniques 
to precisely determine the small effect due to disconnected quark loops.
In the spectrum method, 
we can successfully shift $m_s$ by $\pm 25$~MeV
by using the reweighting technique. 
It would be interesting to study isospin breaking effects,
such as the proton and neutron mass difference, 
by using this technique,
and 
its applicability on larger lattice volumes should be studied.


The ratio method requires precise calculation of 
the nucleon disconnected three-point function,
which is technically very challenging.
The low-lying modes of the Dirac operator turned out to be very helpful: 
we employ the LMA technique to calculate the nucleon propagator 
and the all-to-all quark propagator for the disconnected quark loops.
These techniques, in principle, can be applied to other baryon observables.
For instance, it is interesting to extend this study to 
the strange quark spin content of the nucleon.
Precise knowledge of this quantity is important to 
constrain the parameter space of SUSY models 
through spin-dependent scattering cross section 
of the neutralino-nucleon scattering~\cite{Ellis:2008hf}.

\begin{acknowledgements}
Numerical simulations are performed on 
Hitachi SR11000 and 
IBM System Blue Gene Solution 
at High Energy Accelerator Research Organization (KEK)
under a support of its Large Scale Simulation Program (No.~06-13,
07-16, 08-05, 09-05, 09/10-09 and 10-11) as well as on
NEC SX-8 and Hitachi SR16000 at YITP, Kyoto University and
NEC SX-8R at RCNP, Osaka University.
This work is supported in part by Grant-in-aid for Scientific
Research of Japan(Nos.
~18340075, 
~20105001, 
~20105002, 
~20105005, 
~21674002,
~21684013,
~22224003) and 
Grant-in-Aid for Scientific Research on Innovative
Areas (No.~2004:~20105001,~20105002,~20105003,~20105005) .
\end{acknowledgements}

\end{document}